\documentclass[runningheads]{llncs}
\usepackage{graphicx}
\usepackage{adjustbox}
\usepackage{booktabs}

\begin{document}
\title{Time-position characterization of conflicts: a case study of collaborative editing}
\titlerunning{Time-position characterization of conflicts}
%
\author{Hoai Le Nguyen\inst{1}\orcidID{0000-0002-2978-5249}\newline 
\and
Claudia-Lavinia Ignat \inst{1}\orcidID{0000-0002-0487-6128}}
\authorrunning{H.L. Nguyen and C.-L. Ignat}
%
\institute{ Université de Lorraine, CNRS, Inria, LORIA, F-54000, Nancy, France\\
 \email{\{hoai-le.nguyen,claudia.ignat\}@inria.fr}}

\maketitle              
\begin{abstract}
Collaborative editing (CE) became increasingly common, often compulsory in academia and industry where people work in teams and are distributed across space and time. We aim to study collaborative editing behavior in terms of collaboration patterns users adopt and in terms of a characterisation of conflicts, i.e. edits from different users that occur close in time and position in the document.
The process of a CE can be split into several editing \textit{`sessions'} which are performed by a single author (\textit{`single-authored session'}) or several authors (\textit{`co-authored session'}). This fragmentation process requires a pre-defined `maximum time gap' between sessions which is not yet well defined in previous studies.
In this study, we analysed CE logs of 108 collaboratively edited documents.
We show how to establish a suitable `maximum time gap' to split CE activities into sessions by evaluating the distribution of the time distance between two adjacent sessions.
We studied editing activities inside each \textit{`co-author session'} in order to define potential conflicts in terms of time and position dimensions before they occur in the document. We also analysed how many of these potential conflicts become real conflicts. Findings show that potential conflicting cases are few. However, they are more likely to become real conflicts.

\end{abstract}

\section{Introduction}
\label{intro}

Today, modern word processors such as Google Docs \cite{GoogleDocs}, ShareLaTeX \cite{ShareLaTeX2017}, Etherpad \cite{Etherpad} are popular with many useful features to support collaborative editing (CE) such as adding comments, in-line communication (chat), revision histories and editing logs. The question \textit{`How people write together'\cite{Posner1992,Olson2017}} captured the attention of CSCW researchers. Birnholtz et al \cite{Birnholtz2012} reveal that edits and comments in CE often carry social meaning i.e. they can have emotional and relational impact. The authors also pointed out that communication can be used to explain potentially conflicting behaviors and avoid negative relational effect. Their follow up research \cite{Birnholtz2013} presented an experimental study of group maintenance in collaborative editing using Google Docs. This is the first research that considered to analyse editing logs. The study was separated into an asynchronous phase followed by a synchronous phase. In the asynchronous phase two users separately edited a document on a specific topic and then shared it with the assigned partner who provided feedback and revised the document. The synchronous phase required that the two users edit a shared document summarizing their opinions expressed in their previously written documents. However, the study was controlled by separating the writing activity into asynchronous and synchronous and has not given users the freedom to choose and alternate the writing style. Moreover, the study focused uniquely on the relationship between communication, editing and collaborators social relationships and did not study the editing process of users.

Follow up research studied how people collaboratively edit documents by analysing collaborative editing logs.
Sun et al \cite{Sun2014} presented an analysis of collaboration logs over two years of all Google employees using Google Docs suite. They found that collaboration editing has grown rapidly up to 53\% during the period they examined and `concurrent editing is sticky' with 76\% of the employees who participated in a `concurrent session' repeating the activity in the following month. In \cite{IgnatCDVE14,IgnatECSCW15} authors studied the effect of delay on the error rate, redundancy and quality of collaboratively produced documents by analyzing logs of real-time collaborative editing tasks using Etherpad. Olson et al \cite{Olson2017} examined the traces of collaborative writing behavior of advanced undergraduates in a project course using Google Docs.  They found that  95\% of documents have some simultaneous work (i.e have at least one `co-authored session'). The study assesses the quality of the collaboratively edited documents and analyses different aspects of CE using the taxonomy of CE \cite{Posner1992,Lowry2004}.
D'Angelo et al \cite{DAngelo2018} analysed the histories of a large collection of documents edited in Etherpad\cite{Etherpad} to study how people are writing in the wild and found that simultaneous editing happens very rarely.

Conflict is a common phenomenon in collaboration between groups of
people, and conflict management is a key concern in designing collaborative applications \cite{Easterbrook1993}. In collaborative editing conflicts occur when users concurrently write in the same part of the document. As stated in \cite{DourishCSCW96}, syntactic conflicts occur at the system infrastructure
level, while semantic conflicts are inconsistencies from the perspective of
the application domain. Generally, merging algorithms underlying the application 
solve the syntactic inconsistency problems in collaborative text editing, but
they do not enforce semantic consistency. In \cite{NguyenJCSCW18} authors studied conflicts in asynchronous collaboration over open source software projects that used Git.

We aim to study collaborative editing behavior in terms of patterns of collaboration users adopt such as alternating synchronous and asynchronous collaboration and measuring and comparing user performances during the different collaboration modes. We also aim to study a characterisation of conflicts in terms of time and position in the document.


The process of collaborative editing can be split into several \textit{sessions}, including \textit{single-authored sessions} and \textit{co-authored sessions}. The previous studies did not well define a suitable `interval' or `maximum time gap' which is used for this fragmentation process. Moreover, they haven't provided a detailed analysis of editing activities inside these sessions. We particularly aim to analyse \textit{collaborative edits} inside \textit{co-author sessions} and study how users manage `potential conflict' cases when they edit together in a close period of time and in close parts of the document. For this purpose we define a characterisation of collaborative editing by means of  \textit{time-position windows}. Our research questions are listed below:

\begin{enumerate}
  \item How to choose a suitable `maximum time gap' to split editing activities into sessions?
  \item What is the time-position characterization of editing sessions, namely for `co-authored sessions'?
  \item Inside  `co-authored sessions', how often `potential conflicts' happen within some time-position extension (condition)?
\end{enumerate}

The rest of the paper is organized as follows. In section \ref{related-work}, we present related approaches which are based on an analysis of the traces of collaborative editing. In section \ref{measurements}, we describe the measurements of our study. We then discuss about our results and conclusion of this study.

\section{Related work}
\label{related-work}

Sun et al \cite{Sun2014} published  an in-house study that analysed the logs of activity for all Google employees from 2011 to 2013. They found that on that period, the percentage of new employees who collaborate on Google Docs per month has risen from 70\% to 90\%. To estimate the percentage of documents which had concurrent editing, they used a \textit{15 minutes interval} to split documents into intervals and consider edits by different users in the same \textit{15 minutes intervals} as concurrent edits. The choice of \textit{15 minutes intervals} is arbitrary. And this approach has edge cases in which two users edit the same document within 15 minutes but they are split into two adjacent intervals and are not counted as concurrent edits. Authors proposed a more accurate approach which is looking for a sequence of edits by different users with the maximum gap of 15 minutes. However this proposed mechanism was not applied.

Olson et al \cite{Olson2017} collected and analysed 96 Google Docs documents written by 32 teams of undergraduate students from the Project Management class in three successive years (2011, 2012 and 2013) at University of California, Irvine. They found that 95\% of the documents exhibited some simultaneous work. In fact, they used the approach of \cite{Sun2014} with the \textit{7 minutes gap}. To determine the \textit{7 minutes gap},  they examined all documents with \textit{15 minutes gap} and found that 90\% of them were 7 minutes or less. In more details, a document consists of many \textit{sessions}. Each \textit{session} combines a series of \textit{slices}. Each \textit{slice} which aggregates a series of keystrokes is generated after a certain of pause or a certain amount of edits. If a \textit{session} was edited by more than one editor, they consider it as a \textit{simultaneous session}. The others are considered as \textit{solo-authored sessions}.

Both studies above focus only on time-dimension of collaborative editing. If two authors edit a document within 7 or 15 minutes gaps, they are considered as having a simultaneous writing session either they can edit in adjacent positions or far different positions. The choices of 7 or 15 minutes gaps are still arbitrary. In another study, Larsen et al \cite{Larsen-Ledet2019} use a mixed methods involving interviews and analysis of  the traces of collaborative editing documents (using Google Docs) to outline the role of `territorial functioning' in CE. On their analysis, they take into account the position-dimension of edits to visualize the `editing territories' of different authors over the time. However, for the time-dimension,  they use the same technique as previous studies of Sun et al\cite{Sun2014} and Wang et al \cite{Wang2015} which is based on the \textit{15 minutes time gap}.

D'Angelo et al \cite{DAngelo2018} presented a study on how Etherpad, a real-time collaborative editing tool, is used in the wild. They analysed the histories of a large collection of documents (about 14000 pads) in both time and position dimensions. Edits are independently classified as collaborative or not in time, position and time-position. An edit is considered as collaborative in time dimension if it is \textit{close enough in time to an edit applied by a different author}. In this study, they used the \textit{time windows} which are 5, 10 and 60 seconds to determine if an edit is \textit{close enough} or not. And similarly, they used the \textit{position window} of 10, 80, 400 and 800 characters. For two-dimensional analysis (time-position), all pairs of \textit{time/position windows} were used. Results show that about half of the pads were edited by a single author. Asynchronous collaboration in which users edit in close positions of the document but in different times happens often. Simultaneous editing in which users edit in close positions within the same \textit{time window} happens very rarely. Note that they used the proportion of time, position and time-position collaborative edits over the total edits of the documents for their inferences.

While \cite{Sun2014} and \cite{Olson2017} focus on finding if a document has some \textit{simultaneous sessions} or not, \cite{DAngelo2018} focuses on finding the quantity of \textit{simultaneous edits} of shared editing documents. It presents a more detailed quantitative analysis of \textit{collaborative editing} than the two previous works. However it lacks the overview of how people work together. For example, people can use `divide and conquer' strategy in which editors work in different parts (positions) of the document \cite{Wang2015}. Then it's obviously that the document presents only \textit{time collaborative edits} results on their analysis. Beside, the \textit{5 seconds or 10 seconds time window} is too short to have  multiple editing activities. People can stop to discuss or to read the work of the others during several minutes before continue to write. Moreover, when people are free to collaborate, they do not edit simultaneously all the time. There are several \textit{sessions} that they work asynchronously \cite{Olson2017}.

\section{Time and position characterisation}
\label{measurements}
We analysed \cite{ShareLaTeX2017} logs which were collected from a ShareLaTeX server used inside an engineering school and anonymized for privacy purpose. Groups of three or four students were assigned a writing task and required to use a shared ShareLaTeX document for their collaborative writing. All editing activities inside the shared document were recorded by ShareLaTeX server from the beginning until the end of the assignment, i.e. from 20-September-2017 to 20-November-2017. Students could collaboratively edit the shared documents while being collocated during their classes or remotely from home. However, users were free to use other coordination tools to coordinate their work. We have not analysed their coordination efforts during the task.

In ShareLaTeX, there are two types of edits which are `Insertion' and `Deletion'. They are recorded with the following information: the \textit{timestamp}  when they happen, the \textit{position} in the document where they happen, the \textit{user-id} who performs the edit, the \textit{action-type} which determines the type of edit, i.e. an `Insertion' or a `Deletion' and the \textit{content} which is inserted or deleted. The \textit{content} of an edit can be a single character or a long string. In addition, a copy-paste action is considered as an `Insertion'. A modification action is considered as a `Deletion' of the old content followed by an `Insertion' of the new content.  

We retrieved 1748 documents from the logs. However, 856 documents were created for testing purpose (i.e they were created and edited by a single user and have none or only one edit action). In the rest 892 documents, only 108 of them were edited by more than one author. As we are focusing on collaborative editing, our analysis was performed on these 108 documents. Table  \ref{table:overview-data} presents the overview of our data in which \textit{`No. of authors'} and \textit{`No. of edits'} are the number of editors and the number of recorded editing activities of each documents. The \textit{`Amount of edit'} is the sum of all \textit{content} edits lengths .

\begin{table} [ht]
\begin{center}
\begin{tabular}{@{}lrrrr@{}}
 \toprule
  & \bf Min
  & \bf Max
  & \bf Average
  & \bf Std \\
 \midrule 
\bf No. of authors   &  2 & 4  & 2.69  & 0.87  \\
\bf No. of edits & 53  & 38,329  &  8,000  & 10,583    \\
\bf Amount of edit	& 245  & 272,935 &  47,133    &  56,866 \\
  \bottomrule
\end{tabular}
\end{center}
\caption{Overview of the data: 108 documents}
\label{table:overview-data}
\end{table}

\begin{figure}[ht]
 \centering
  \includegraphics[width=0.8\columnwidth]{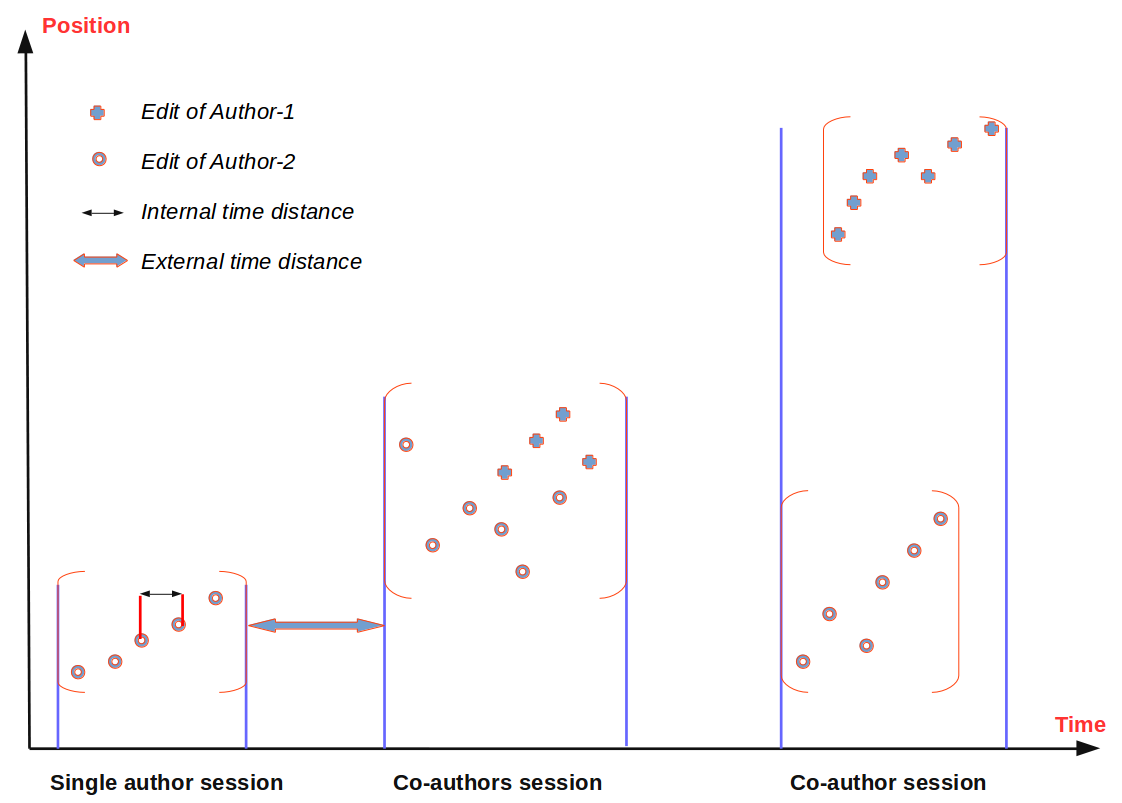}
  \caption{A document with two authors in time-position view}
  \label{fig:document-time-position}
\end{figure}

A document can be presented in time-position view (two dimensional view). Figure \ref{fig:document-time-position} presents a sample document which is segmented into three \textit{writing sessions} by time dimension. These sessions are classified into \textit{single-author-session} (SAS) and \textit{co-author-session} (CAS) depending on the number of editors of each session. In this sample we have one SAS and two CASs. Note that in a CAS, two or more editors can edit in the same position or in different positions. For a \textit{time dimension} analysis we defined `internal time distance' (or \textit{internal-distance}) which is the time distance between two adjacent edits in the same session and `external time distance' (or \textit{external-distance}) which is the time distance between two adjacent sessions in a document.

\subsection{Time dimension}
\label{time-dimension}
We first borrow the proposed approach of \cite{Sun2014} to analyse the `time dimension' of our data. Instead of using only an \textit{`arbitrary maximum time gaps'}, we try to examine the data with different \textit{`maximum time gaps'} : 15 minutes, 7 minutes, 5 minutes, 2 minutes, 1 minute and 30 seconds. Furthermore, after dividing a document into sessions and classifying the sessions into SASs and CASs, we analysed the differences between CASs and SASs such as: the internal-distance which is the distance between two edits in the same session, the average time which is the average length of sessions, the average number of edits of sessions.

\begin{table}
\begin{center}
\begin{adjustbox}{max width=.92\textwidth}
\begin{tabular}{@{}lrrrrrr@{}}
 \toprule
  & \multicolumn{6}{c}{\bf Time gaps} \\
  & 15mn & 7mn & 5mn & 2mn & 1mn & 30s\\
\midrule
\textbf{Doc having CASs} & 84/108 & 82/108 & 80/108 & 77/108 & 76/108 &75/108\\
  Proportion & 77.77\%  & 75.92\%  & 74.07\%  & 71.30\%  & 70.00\% & 69.44\% \\[.5em]
\textbf{No. of CASs per doc} \\
  Average  & 2.3  & 2.9  & 3.4  & 5.8  & 9.4 & 14.1\\
  Proportion & 28.4\% & 24.4\% & 22.7\%  & 17.5\% & 13.9\% & 10.7\%\\[.5em]
\textbf{Internal-distance} \\
  SASs (Average)  & 9.61s& 6.77s  & 5.89s  & 4.04s  & 2.87s & 2.01s\\
  SASs (CI 99\%)  & [6.56-12.68]& [5.74-7.80]  & [5.19-6.61]  & [3.73-4.35]  & [2.74-3.00] & [1.97-2.07]\\
  CASs (Average) & 4.25s  & 4.19s  & 4.15s  & 2.55s & 1.78s & 1.24s\\
  CASs (CI 99\%) & [2.89-5.62]  & [2.69-5.70]  & [2.71-5.59]  & [1.97-3.14]&[1.54-2.02] & [1.13-1.35]\\[.5em]
\textbf{Session length}  &    \\
 SASs (Average)     & 972s & 647s & 507s & 226s & 109s & 51s\\
 CASs (Average)    & 3,314s & 2,369s & 1,953s & 878s & 350s & 155s \\[.5em] 
\textbf{No. of edits} &\\
 SASs (Average)  & 213 & 170 & 146 & 94 & 60 & 39\\
 CASs (Average)  & 2,140 & 1,841 & 1,629 & 961 & 470 & 255 \\
 CASs (Normalized)& 893 & 787 & 704 & 429 & 214 & 118\\
\bottomrule
\end{tabular}
\end{adjustbox}
\end{center}
\caption{Documents segmentation by different \textit{maximum time gaps}}
\label{table:time-segmentation}
\end{table}

Table \ref{table:time-segmentation} presents all the results of our analysis in time dimension. \textit{Doc having CAS(s)} shows the number of documents that have at least one co-author session. The proportion of  \textit{Doc having CAS(s)} over all analysed (108) documents is 77.77\% with \textit{ 15 minutes time gap} and downs to 69.44\% with \textit{ 30 seconds time gap}. In comparison to \cite{Olson2017} which showed that  95\% of documents exhibited some `simultaneous work' with \textit{7 minutes time gap}, our data set shows that 75.92\% of documents have collaborative sessions. Also with this time gap, our analysis shows that the average length of co-author sessions is 2369 seconds (39.5 minutes) and the longest co-author session is 8639 seconds (144 minutes) while they are 9.2 minutes (average) and 74 minutes (longest) in \cite{Olson2017}.

\textit{No. of CASs per Doc} is the average number of CASs in each document after a segmentation of the document with the given \textit{time gaps}. We also displayed the proportion of CASs over all sessions (CASs and SASs). The smaller \textit{time gaps} given, the more CASs are generated. However, it reduces the proportion of CASs over the total sessions. In another way, the number of CASs increases slower than the number of SASs when the \textit{time gap} decreases. 

\textit{Internal-distance} presents the average internal distance between two edits in the same session (SASs or CASs). For more details, we calculated the confidence interval with 99\% of significance (CI 99\%) for the internal distance variable. We found that the internal distance of SASs is longer than the internal distance of CASs. In another way, the distance between two edits in single-author sessions is longer than the one in co-authors sessions. 

\textit{Session length} shows the average length in seconds of each session, i.e. how long each session lasts. \textit{No. of edits} is the average number of edits in each session. We found that (with 99\% of significance) the average length of CASs is longer than  the average length of SASs and also that CASs have larger number of edits than SASs in average. In order to compare \textit{No. of edits} for CASs and SASs, we normalized \textit{No. of edits} of CASs as it includes edits from all collaborators while \textit{No. of edits} of SASs includes only edits of a single editor. Normalized \textit{No. of edits} of each document is calculated by  dividing original \textit{No. of edits} to the number of collaborators.  Having more edits with shorter distance between edits and longer collaborating time, it significantly gives us a quantitative view that the co-authors sessions are more productive in terms of the quantity of contributions to the documents than the single-author sessions.

\begin{figure}
 \centering
  \includegraphics[width=0.8\columnwidth]{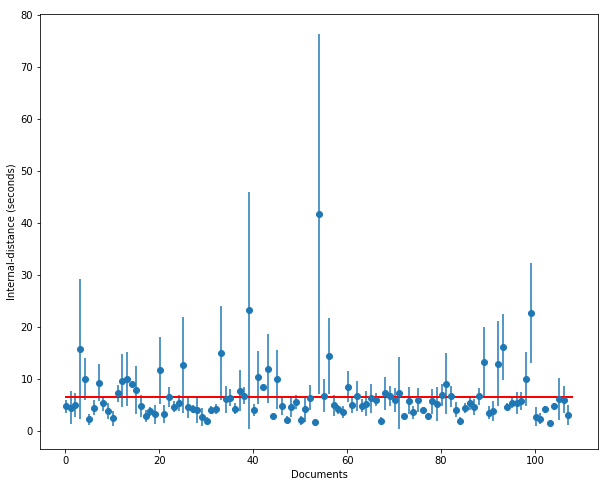}
  \caption{Time gap = 420s, Average internal-distance with confidence interval CI90}
  \label{fig:internal-420-CI90}
\end{figure}

As the \textit{maximum internal-distance} is limited by the \textit{time gap}, it's obviously that the \textit{internal-distance} presented in Table \ref{table:time-segmentation} must be shorter than the given \textit{time gap}. However, the result shows that the average \textit{internal-distance} is much shorter than expected. For a better understanding, we calculated the confidence interval of \textit{internal-distance} for each documents with 90\% significance. Both SASs and CASs were included in this analysis. Figure \ref{fig:internal-420-CI90} presents the results of \textit{420s (7 minutes) time gap} in which all the confidence intervals do not reach out 80 seconds and the general average \textit{internal-distance} of all documents is 6.5 second presented by the red line. We can see that the \textit{7 minutes time gap} is not a suitable \textit{time gap} for our corpus. The suitable \textit{time gap} should be shorter than 7 minutes.

\begin{figure}
 \centering
  \includegraphics[width=0.8\columnwidth]{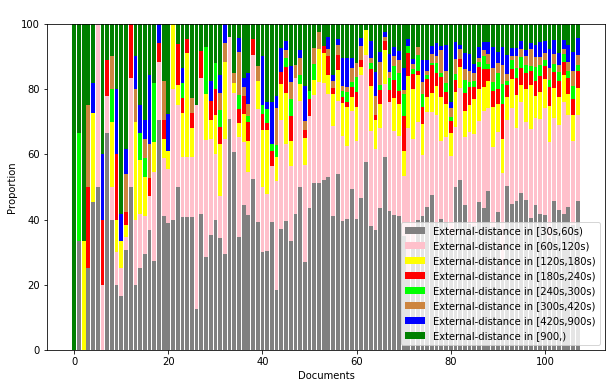}
  \caption{Time gap = 30s, External-distances distribution}
  \label{fig:timegap30-external-distribution}
\end{figure}

In addition, we measured the \textit{external-distances} of \textit{30 seconds time gap} and calculated their distribution in different intervals which are created from \textit{potential time gaps}: [30s,60s) , [60s,120s), [120s,180s), [180s,240s), [240s,300s), [300s,420s), [420s,900s), [900s,). In which the left square bracket `[' denotes  \textit{`equal or longer than'}, the right round bracket `)' denotes \textit{`shorter than'} and the last interval denotes the \textit{external-distances} which are \textit{`equal or longer than 900 seconds'}. Our suggestion is that if an interval covers more \textit{external-distances} than others, it has much `potential' to become a suitable \textit{time gap} than others. Figure \ref{fig:timegap30-external-distribution} shows the distribution of \textit{external-distances} in each document. In average, \textit{external-distances} represent 39.73 \%, 24.58\%, 8.98\%, 4.37\%, 2.53\%, 3.03\%, 3.98\% and 12.80\% respectively for the given intervals. This means that if we increase the \textit{time gap} from 30 seconds to 60 seconds, 39.73\% of sessions will become part of other sessions because \textit{external-distances} can not be shorter than 60 seconds. If we use 120 seconds \textit{time gap} instead of 30 seconds \textit{time gap}, 64,31 \% (39.73\% +24.58\%) of sessions will be merged into other sessions and so on. From the above results, we can say that the [30s, 60s) and [60s,120s) intervals have much potential to contain the suitable \textit{time gap} as it covers much more \textit{external-distances}(64,31\%) than others. Moreover, we found that the range of \textit{external-distances} is very wide, from 30 seconds to 87 hours (~3.6 days). This wide range is due to the fact that collected logs represent the students writing task over a period of two months with an allocated time slot in their schedule of two hours per week, but students could continue the writing outside the allocated time slot.


Summarizing the time-dimension analysis, we found that collaborative editing is usually separated into many editing sessions including \textit{single-author sessions} and \textit{co-author sessions}. The time distance between sessions has a very wide range (up to 87 hours in our case study). To split a document into sessions, a suitable \textit{time gap} needs to be determined. And finally, editors have more editing activities in co-authors sessions than in single-author sessions, i.e. having more contributions when working collaboratively.

\subsection{Time-position analysis}
The analysis on \textit{time dimension} gives us an overview of how collaborative editing happens over the time. It determines \textit{co-authors sessions} in which the authors write closely together in time. However, it lacks the information about whether or not they write `closely' in the same part of a document or `separately' in different parts of it. A more detailed analysis in both time-position dimensions gives us a better understanding about how they write collaboratively.

\begin{figure}
 \centering
  \includegraphics[width=0.8\columnwidth]{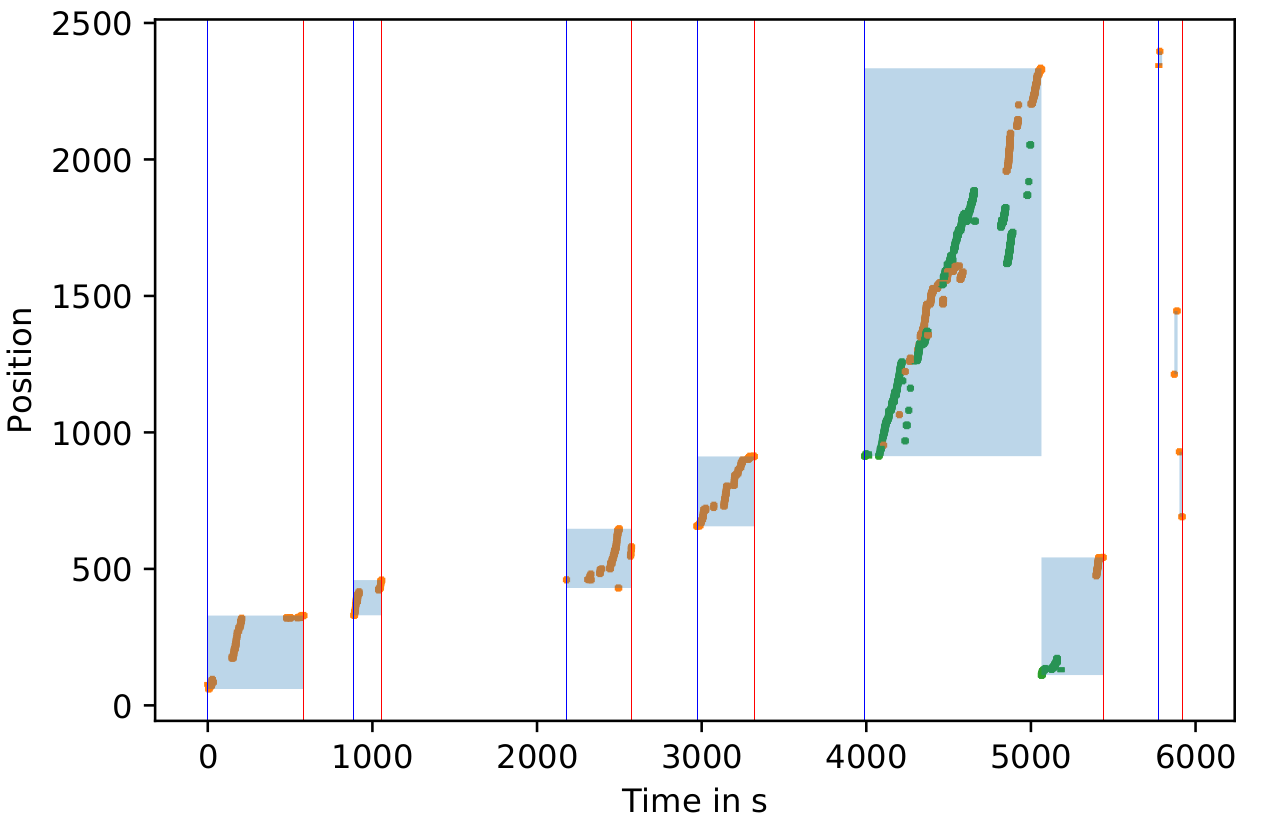}
  \caption{A real document(id=59e8f8d98e96ef7e2dc01eb2) presented in time-position view}
  \label{fig:real-document}
\end{figure}

As  we can see in Figure \ref{fig:document-time-position}, edits inside a session can be grouped into different `clusters' depending on their time-position distances. In order to explain our time-position analysis we illustrate in Figure \ref{fig:real-document} the time-position view of the document with id \textit{59e8f8d98e96ef7e2dc01eb2} from our data corpus.  The figure illustrates the notions of session and cluster. A session begins with a vertical blue line and ends with a  vertical red line. A cluster is presented as a rectangle filled with light-blue color. Edits are presented as orange and green dots depending on which authors they belong to (called orange author and green author). This document contains about 2500 characters (including white-spaces, empty lines) and was edited in total 6000 seconds by 2246 edit actions (including deletion, insertion). In the first 60 minutes, it was edited by the orange author only. Those edits are split into four single-author sessions, the longest time distance between them being about 16 minutes. In the next 25 minutes, the figure illustrates a co-authors session in which the document was edited collaboratively by two authors (the orange author and the green author). And in the last time slot, the figure illustrates a single-author session of the orange author. In this session, the orange author had edited in three different positions of the document with position-distance larger than 400 characters. Note that in Figure \ref{fig:real-document}, for the simple presentation, we use large size windows of the form \textit{[time-gap, position-gap] = [300seconds, 400 characters]} in order to reduce the number of sessions and clusters.

Edits in a single-author session can be re-edited by another author in another session. However, these two sessions are separated in time so that if conflicts happen, they are asynchronous conflicts. In this analysis, we focus on the cases that two or more authors edit closely together in both time and position. Having a closer look in the co-author session in Figure \ref{fig:real-document} which contains two clusters of edits, we can see that these two clusters have edits of both authors. In the bottom right cluster, there is a clear border between edits of two authors. In the top left cluster, besides three borders separating edits of different authors, there are several cases in which one author edited between two continuous edits of another author. We are interested in a characterization of conflicts in these cases where all involved edits are very close together in both time-position dimensions. 

\begin{figure}
 \centering
  \includegraphics[width=0.8\columnwidth]{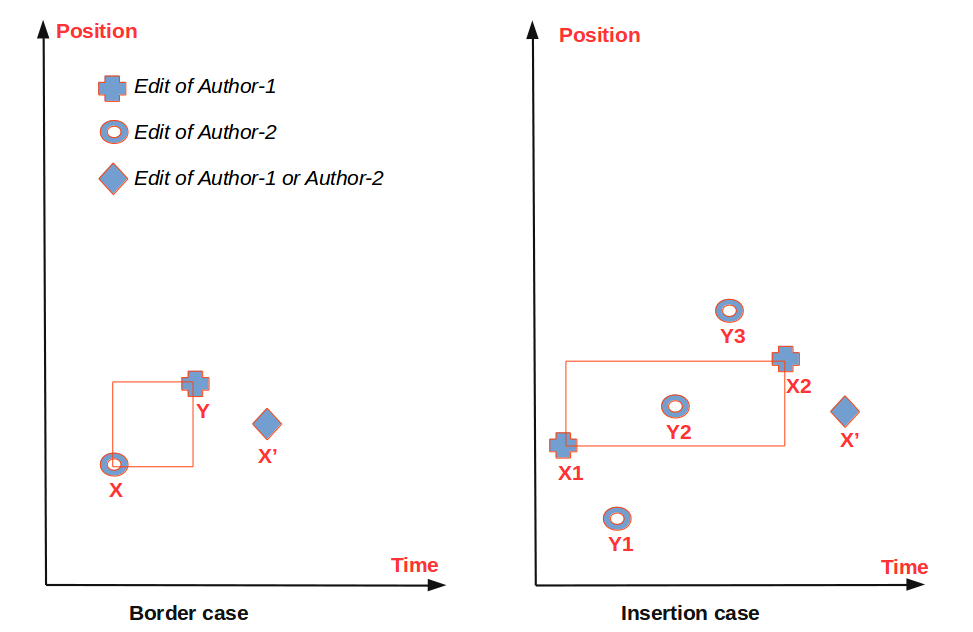}
  \caption{Illustration of Border case and Insertion case}
  \label{fig:border-insertion-case}
\end{figure}

We examined two cases in which conflicts potentially happen. The first case called \textit{border case} refers to the switch-point between two adjacent editing-areas which belong to two different authors. Those editing-areas can contain one or more continuous edits. The second case called \textit{insertion case} refers to the case in which one author tries to edit between two continuous edits of another author,i.e. one author tries to insert one or more edits into an editing-area of another author.  In both cases, if the time-position distance between those continuous edits is small, conflicts have high potential to happen. Figure \ref{fig:border-insertion-case} presents the illustration of \textit{border case}  and \textit{insertion case}. On the left side of Figure \ref{fig:border-insertion-case}, \textbf{X, Y} and \textbf{X'} are three continuous edits (in time order) of two different authors in which \textbf{[X, Y]} form a \textit{border case}. This \textit{border case} can become a \textit{potential border conflict case} if its time-position distance defined by the red rectangle is small. And if the adjacent edit  \textbf{X'} happens between the positions of \textbf{X} and \textbf{Y}, it should be classified as a \textit{border conflict}. A formal description of \textit{border conflict} is presented in Definition \ref{def:border-conflict}.

On the right side of Figure \ref{fig:border-insertion-case}, [$X_1, Y_1, Y_2, Y_3, X_2, X'$] is a sequence of edits in time order. It means that $Y_1, Y_2, Y_3$ happen between $X_1$ and $X_2$ and $X'$ happens right after $X_2$. However, in position dimension, the order is different, which is [$Y_1$, $X_1$, $Y_2$, $X'$, $X_2$, $Y3$]. Focusing on [$X_1, Y_2, X_2$], we can see that they satisfy both time order and position order. In another way, \textit{Author-2} had inserted an edit into the time-position window formed by two continuous edits of \textit{Author-1}. In the Figure \ref{fig:border-insertion-case}, this time-position window is presented by the red rectangle created by $X_1$ and $X_2$. If this window is small and the adjacent edit \textbf{\textit{X'}} happens between the position of $X_1$ and $X_2$,we consider this case as an \textit{insertion conflict}. A formal description of \textit{insertion conflict} is presented in Definition \ref{def:insertion-conflict}.
 
\begin{definition}
\textit{A sequence of edits [$X_1, X_2, .., X_n$] is a \textbf{sequence of edits in time order with time-gap t} if } $\forall{i} \in [1,n): time(X_{i+1}) > time(X_i) $ and \textit{time-distance}$(X_i, X_{i+1}) < t$
\label{def:sequence-time-order}
\end{definition}

\begin{definition}
\textit{A sequence of edits [$X_1, X_2, .., X_n$] is a \textbf{sequence of edits in position order with position-gap p} if } $\forall{i}\in [1,n): position(X_{i+1}) > position(X_i)$ and \textit{position-distance}$(X_i, X_{i+1}) < p$
\label{def:sequence-position-order}
\end{definition}

\begin{definition}
\textit{X is an edit of \textit{Author-1}, Y is an edit of \textit{Author-2} and X' is an edit belonging to one of them. If [$X, Y, X'$] is a sequence of edits in time order with time-gap t and  [$X, X', Y$] is a sequence of edits in position order with position p, [X,Y] then form a \textbf{border conflict} within time-position window [t,p].}
\label{def:border-conflict}
\end{definition}

\begin{definition}
\textit{$X_1, X_2$ are edits of \textit{Author-1}, $Y_1, Y_2, ... Y_k$ are edits of \textit{Author-2} and X' is an edit belonging to one of them. $(Y_i)^+, i \in [1,k]$ is a sub-sequence of edits of \textit{Author-2} which has at least one edit. If [$X_1, Y_1, Y_2,  ... Y_k, X_2, X'$] is a sequence of edits in time order with time-gap t and $\exists{(Y_i)^+}$ so that [$X_1$, $(Y_i)^+$, $X_2$] or [ $X_2$, $(Y_i)^+$, $X_1$] form a sequence of edits in position order with position p, [$X_1$, $(Y_i)^+$, $X_2$] or [$X_2$, $(Y_i)^+$, $X_1$) then form an \textbf{insertion conflict} within time-position window [t,p].}
\label{def:insertion-conflict}
\end{definition}

We use a \textbf{[30s, 10c]} time-position window to run our experiments. As we explained in Section \ref{time-dimension}, the documents are separated into sessions using a \textit{30 seconds time-gap}. After that, all co-authors-sessions are checked for \textit{border cases} and \textit{insertion cases}. If these \textit{border cases} and \textit{insertion cases} satisfy the selected time-position window which is \textbf{[30s, 10c]}, they become \textit{potential border conflicts} and \textit{potential insertion conflicts}. And if one of the involved authors edits right after in the potential-conflict-area, we consider that potential conflict as a conflict.
The reason that we choose the  \textbf{[30s, 10c]} time-position window is that it can allow three or more editing actions and can cover the position distance of two or three words. Table \ref{table:30s-10c-window} characterizes conflicts for the \textbf{[30s, 10c]} time-position window.

\begin{table}[ht]
\begin{center}
\begin{tabular}{@{}lrrr@{}}
 \toprule
  & {\textbf{Border conflict}}
  & \phantom{ab}
  & {\textbf{Insertion conflict}} \\
  \midrule
Proportion of Potential-conflicts  &  5.7\% && 2.27\%  \\
over Consider-cases CI99\% & [1.73-9.66\%] && [0-5.04\%]\\[.5em]
Proportion of Conflicts & 84.52\% && 97.22\%   \\
over Potential-conflict CI99\% &[77.53-91.51\%] && [88.96-100\%]\\[.5em]
Average of Time-distance &6.17s && 4.06s \\
of Conflict cases CI99\% & [3.67-8.68s] && [0-10.3s]\\[.5em]
Average of Position-distance  &3.43c && 4.15c \\
of Conflict cases CI99\% & [2.88-3.98c] && [2.13-6.17c]\\
\bottomrule
\end{tabular}
\end{center}
\caption{Border conflict and Insertion conflict with \textbf{[30s, 10c]} time-position window}
\label{table:30s-10c-window}
\end{table}

The results in Table \ref{table:30s-10c-window} show the high proportion of potential-conflicts that become conflicts: from 77.53\% to 91.51\% for \textit{border conflict} and from 88.96\% to 100\% for \textit{insertion conflict} with significance of CI99\%. However, these two types of conflict happen very rarely. It is less than 9.66\% for \textit{border conflict} and less than 5.04\% for \textit{insertion conflict}. The case of \textit{potential border conflict} that is not a \textit{border conflict}  corresponds to the case that the time-position window (the border) of two continuous edits of two authors is larger than the time-position window that we use to determine \textit{border conflict}. And in the case of \textit{potential insertion conflict} that is not an \textit{insertion conflict}, two authors are editing in two different areas which are large enough in position.

Beside the time-position window \textit{[30s, 10c]}, we also used a smaller window of \textit{[10s, 5c]} and a larger window of \textit{[60s, 20c]} to examine the \textit{border conflict} and the \textit{insertion conflict}. Results are presented in Table \ref{table:10s-5c-window} and Table \ref{table:60s-20c-window} respectively. We can see that the \textit{potential border conflict}  is affected by the time-position window more than the \textit{potential insertion conflict}. The \textit{potential border conflict} decreases from 5.7\% to 3.07\%  with a smaller window and increases to 9.94\% with a larger window. The \textit{potential insertion conflict}  has less effects by the size of time-position window. Furthermore, the smaller time-position window decreases the proportion of \textit{border conflict} over \textit{potential border conflict} while the larger window increases it. For the \textit{insertion conflict}, the result is reversed. It means that the smaller the time-position window is, the more likely the \textit{potential insertion conflicts} become real \textit{insertion conflicts}.

\begin{table}[ht]
\begin{center}
\begin{tabular}{@{}lrrr@{}}
 \toprule
  & {\textbf{Border conflict}}
  & \phantom{ab}
  & {\textbf{Insertion conflict}} \\
  \midrule
Proportion of Potential-conflicts  & 3.07\% && 2.13\%  \\
over Consider-cases CI99\% & [0.9-5.23\%]&& [0-5.49\%]\\[.5em]
Proportion of Conflicts & 84.04\% && 100\%   \\
over Potential-conflict CI99\% &[76.93-91.16\%] && [NA]\\[.5em]
Average of Time-distance &2.59s && 4.34s \\
of Conflict cases CI99\% & [1.51-3.68s] && [0-12.91s]\\[.5em]
Average of Position-distance  &2.23c && 2.3c \\
of Conflict cases CI99\% & [1.88-2.57c]&& [0.98-3,62c]\\
\bottomrule
\end{tabular}
\end{center}
\caption{Border conflict and Insertion conflict with \textbf{[10s, 5c]} time-position window}
\label{table:10s-5c-window}
\end{table}

\begin{table}[ht]
\begin{center}
\begin{tabular}{@{}lrrr@{}}
 \toprule
  & {\textbf{Border conflict}}
  & \phantom{ab}
  & {\textbf{Insertion conflict}} \\
  \midrule
Proportion of Potential-conflicts &9.94\% && 2.04\%  \\
over Consider-cases CI99\% & [3.95-15.93\%] && [0-4.33\%]\\[.5em]
Proportion of Conflicts & 87.0\% && 95.53\%   \\
over Potential-conflict CI99\% &[80.98-93.02\%] && [85.05-100\%]\\[.5em]
Average of Time-distance &5.33s && 4.24s \\
of Conflict cases CI99\% & [3.16-7.5s] && [0-9.14s]\\[.5em]
Average of Position-distance  &7.46c && 4.9c \\
of Conflict cases CI99\% & [5.96-8.97c] && [2.41-7.39c]\\
\bottomrule
\end{tabular}
\end{center}
\caption{Border conflict and Insertion conflict with \textbf{[60s, 20c]} time-position window}
\label{table:60s-20c-window}
\end{table}

As an implication for design for our study we recommend that awareness mechanisms \cite{DourishCSCW92} could be proposed for users when potential conflicts of both types border conflicts and insertion conflicts are detected. Users can get notified
by means of a `heat map' that visualizes the recency of editing activities \cite{Larsen-Ledet2019} when they write closely in time and position with other users, i.e. when the potential border conflicts and potential insertion conflicts occur.

\section{Conclusion}
\label{conclusion}
In this paper we studied collaborative editing behavior in terms of collaboration patterns users adopt and in terms of a characterisation of conflicts, i.e. edits from different users that occur close in time and position in the document. By examining different `maximum time gaps' from 30 seconds to 15 minutes we found that the time distance between sessions (i.e \textit{`external-distance'}) has a very wide range (up to 87 hours in our case study). By evaluating the distribution of \textit{`external-distance'} of a very small time gap, we can determinate a suitable `maximum time gap' to split editing activities into single-author sessions and co-author sessions. We found that users are more productive in co-author sessions than in single-author sessions. 

In a more detailed analysis of the \textit{co-authors-sessions}, we use a [30 seconds, 10 characters] time-position window to examine the cases in which two authors edit closely together in both time and position. We focus on two cases which potentially result in conflict: \textit{`border case'} and \textit{`insertion case'}. \textit{`Border case'} refers  the cases in which two different authors edit in the border of two close editing areas that belong to them. And \textit{`insertion cases'} refers to the cases in which one author does some edits between two continuous edits of another author.  The results show that these two cases happen rarely: up to 5.04\% for \textit{`insertion cases'} and up to 9.66\% for \textit{`border cases'}. 
It means that people rarely edit closely in both time and position. However, these cases (i.e. the case in which people edit closely) are very likely to become conflicts: 77.53\% to 91.51\% of \textit{`border cases'}  and 88.96\% to 100\% for \textit{`insertion cases'} result in `conflict'. From above results, we suggest that collaborative editing tools (ShareLaTeX in this case) should consider to have an awareness mechanism for these two types of `potential conflicts'.

\section*{Acknowledgement}
We would like to thank Gérald Oster and Quentin Laporte-Chabasse for their valuable help for the collection of the ShareLaTeX logs. 
%
%
%
\bibliographystyle{splncs04}
\bibliography{references}






\end{document}